\documentclass[pra,twocolumn,floatfix]{revtex4}
\usepackage{graphicx}
\DeclareUnicodeCharacter{2009}{\,}
\usepackage{color}
\usepackage{amsmath}
\begin{document}

\title{Yrast states of quantum droplets confined in a ring potential}

\author{M. \"{O}gren$^{1,2}$ and G. M. Kavoulakis$^{2,3}$}
\affiliation{$^1$School of Science and Technology, \"{O}rebro University, 70182 
\"{O}rebro, Sweden
\\
$^2$HMU Research Center, Institute of Emerging Technologies, GR-71004, Heraklion, 
Greece
\\
$^3$Department of Mechanical Engineering, Hellenic Mediterranean University, 
P.O. Box 1939, GR-71004, Heraklion, Greece
}
\date{\today}

\begin{abstract}

We consider a quantum droplet which is confined in a ring potential. We investigate 
the so-called ``yrast" state, i.e., the lowest-energy state of the droplet assuming 
that it has some fixed expectation value of the angular momentum. Two are the most 
interesting aspects of this problem, the nonlinear term -- which is partly attractive 
and partly repulsive -- and the periodic boundary conditions. For some range of the 
parameters, the attractive, or the repulsive part of the nonlinear term dominates 
and one gets the expected behavior. In some intermediate regime the two nonlinear 
terms are of comparable size. In this case both the solution, as well as the 
corresponding dispersion relation show an interesting behavior. Finally, we make 
contact with the problem of solitary-wave excitation, since the derived solutions 
are travelling-wave, i.e., solitary-wave, solutions.

\end{abstract}

\maketitle

\section{Introduction}

Quantum droplets are self-bound states, which were first predicted 
by Petrov \cite{Petrov}, in mixtures of Bose-Einstein condensed gases. 
The basic idea behind these quantum objects is the following. In the vast 
majority of experiments on cold-atomic systems these are dilute, and as a
result the mean-field energy is the dominant part of the energy, while the 
corrections \cite{LHY} are very small. However, in a two-component system, 
if the inter- and intra-species coupling constants are tuned, the mean-field 
energy may take any value, or even vanish. In this case, the contribution to 
the energy due to the beyond-mean-field effects is no longer negligible, but 
rather it balances the one due to the mean-field. This is precisely the 
mechanism that gives rise to these self-bound states.

The literature on this problem is very extensive, so here we refer to just 
a few relevant studies, see, e.g., the review articles \cite{rrev1, rrev2}, 
and Refs.\,\cite{PA, th0, th1, th2, th3, th4, th5, th6, th7, th8, th9, th10, 
th11, th12, th14, th15, th16, EK, add0, add1, th166, Lila, NKO, add2, add3, 
add4, add5, add6}. Experimentally, quantum droplets have been observed in 
two-component Bose-Einstein condensed gases \cite{qd7, qd8, qd8a, gd8b, qd8c, 
qd8d}, but also in single-component gases with strong dipolar interactions 
\cite{qd1, qd2, qd3, qd4, qd5, qd6}.

One of the novel aspects of quantum droplets is the presence of an attractive 
and of a repulsive nonlinear term in the generalized Gross-Pitaevskii equation 
that dictates the order parameter. Actually, it is the balance between these 
two terms which makes quantum droplets self-bound. Although the existence of 
a droplet does not require the presence of any trapping potential, clearly 
confining it in some potential may give rise to interesting effects. 

Motivated by this, we study in the present paper the ground-state properties, 
and the rotational properties of a quantum droplet, which is confined in a ring 
potential -- we should stress at this point that such trapping potentials have 
been designed experimentally, see, e.g., Ref.\,\cite{ring}. More specifically, 
we study the so-called ``yrast" problem, which is a terminology used mostly in 
the field of nuclear physics \cite{BM}. Within the yrast problem one is 
interested in the lowest-energy state of the system, for some given value 
of the angular momentum. 

It has been shown that the yrast states are travelling-wave solutions \cite{JKSM}. 
In other words, the problem that we study here is equivalent to the one of 
solitary-wave excitation. Our study is thus also relevant with Refs.\,\cite{ME,
POL}. Reference \cite{ME} identified the criteria for the existence of ``dark" 
solitary waves in quantum droplets. Also, Ref.\,\cite{POL} examined the same 
problem and demonstrated the presence of two families of solitary-wave solutions. 

In what follows below we thus examine the yrast state of a quantum droplet, in a 
purely one-dimensional trapping potential, considering the so-called ``symmetric" 
problem (which is analysed below). One may wonder why we do not 
focus on the more general, asymmetric, problem; actually, this problem was 
considered in Ref.\,\cite{Steffi}. The answer to this question is two-fold. 
First of all, although very interesting -- the ``asymmetric" problem is different 
than the one we consider here. The main reason is that, at least for a substantial 
population imbalance, the droplet coexists with a uniform background, since the 
excess atoms in the majority component can not bind to it. Furthermore, the case 
of a zero (i.e., very small in an actual experiment) population imbalance is 
experimentally relevant.

From the analysis that is presented in the next sections, we distinguish between 
three regimes, depending on the value of the density of the homogeneous state $n_0 
= N/L$, with $N$ being the atom number and $L = 2 \pi R$ the circumference of the 
ring of radius $R$. The three regimes are characterized by a ``low", a ``medium" 
and a ``high" value of $n_0$. In the two opposite limits of low/high density, it 
is the attractive/repulsive nonlinear term that is the dominant one. Reference 
\cite{Lila} examined the same transition using exact diagonalization for a system 
of few atoms. In the intermediate regime, the two terms are comparable and -- clearly 
-- this is the most interesting of the three regimes. 

Furthermore, there are two length scales in the problem, namely the circumference 
of the ring $L$, and the ``natural" size of the droplet $d$ in free space. We focus 
on the regime where $L$ is larger, or comparable with $d$. In the opposite limit 
the density distribution is always homogeneous, which makes the problem analogous
to that of a scalar condensate with an effective repulsive interaction.

In addition, since we work in a ring potential, we impose periodic boundary 
conditions. The problem of solitary-wave solutions in a ring has been studied 
in Refs.\,\cite{Carr1, Carr2, SMKJ} in the case of a scalar condensate with an 
effective contact interaction. The thermodynamic limit is achieved when $N$ 
and $L$ tend to infinity, with the ratio $N/L$ being finite. Interesting effects 
show up in rings of a finite size. Similar conclusions also hold in the present 
problem, for the case of ``medium" and ``high" values of $n_0$. 

Starting with the non-rotating problem, for some parameter range we find that 
there may be two competing solutions, i.e., the homogeneous and localized solutions. 
These two different kinds of solutions may exist simultaneously, with the one being 
the absolute minimum of the energy and the other one being locally stable, i.e., 
metastable. As we give angular momentum to the system, these two solutions may 
become unstable, or their energies may cross. As a result, the yrast spectrum has 
an interesting structure.

In Sec.\,II we present the model and the nonlinear equation that governs the order 
parameter of the system. Then, in Sec.\,III we examine the static problem, starting 
with the local stability of the homogeneous solution of the non-rotating droplet, 
which allows us to derive the phase diagram which separates the homogeneous from 
the localized solutions. In the same section we present the Thomas-Fermi limit of 
our problem. In Sec.\,IV we turn to the problem of rotation, examining this problem 
as the angular momentum is varied, for some representative values of the atom number 
and of the radius of the ring. We first consider the two limiting cases of ``low" 
and ``high" densities, where the derived results are analogous to the ones known 
from previous studies, since the nonlinear term is dominated either by the attractive, 
or the repulsive nonlinear terms. We then turn to the case of ``medium" density,
where the homogeneous phase competes in energy with the localized, which is actually
the most interesting regime. In Sec.\,V we present the connection between our 
derived solutions and the solitary-wave solutions. Finally, in Sec.\,VI we 
present a summary of the main results and an overview.

\section{Model}

As we mentioned above, the problem that we have in mind is that of a quantum droplet 
which is confined in a ring potential of length $L = 2 \pi R$. Therefore, we assume 
purely one-dimensional motion, imposing periodic boundary conditions. The assumption 
of one-dimensional motion is realized under the presence of a harmonic trapping 
transversely to the axis of the motion of the atoms, with an oscillator length 
$a_{\perp}$ which is much smaller than the healing length $\xi$. 

In the symmetric case that we consider, we assume equal masses $M$ for the two components, 
equal populations and also equal coupling constants for the same species $g_{\uparrow 
\uparrow}$ and $g_{\downarrow \downarrow}$, $g_{\uparrow \uparrow} = g_{\downarrow 
\downarrow} = g$. Here $g_{ij}$ result from the tree-dimensional scattering lengths
$a_{ij}$ integrating over the transverse direction and therefore $g_{ij} \sim 
\hbar^2 a_{ij}/(M a_{\perp}^2)$, with $i,j = \uparrow$ and $\downarrow$.

In the symmetric case the two coupled generalized Gross-Pitaevskii equations for the 
two order parameters reduce to a single equation. It was shown in Ref.\,\cite{PA} that 
the (common) order parameter for the two components $\Phi(x,t)$ satisfies the equation
\begin{eqnarray}
  i \hbar \frac {\partial \Phi} {\partial t} = 
  - \frac {\hbar^2} {2 M} \frac {\partial^2 \Phi} {\partial x^2}
  + \delta g |\Phi|^2 \Phi - \frac {\sqrt{2 M}} {\pi \hbar} g^{3/2} |\Phi| \Phi,
  \label{1steq}
\end{eqnarray}
where $\int |\Phi|^2 \, dx = N$, with $-L/2 < x \le L/2$. Obviously, the periodicity 
condition implies that $\Phi(x-L/2,t) = \Phi(x+L/2,t)$. In the equation above 
$\delta g = g_{\uparrow \downarrow} + g$, where $g_{\uparrow \downarrow}$ is the 
coupling constant between the different components, with $0 < \delta g \ll g$ \cite{PA}.
Introducing the units of time $t_0 = \pi^2 \hbar^3 \delta g/(2 M g^3)$, of length $x_0 = 
\hbar^2 \pi \sqrt{\delta g}/(M \sqrt{2 g^3})$, and $\Phi_0 = \sqrt{2 M g^3}/(\pi \hbar 
\delta g)$, we can write for the dimensionless order parameter $\tilde{\Phi} = \Phi/\Phi_0$ 
\begin{eqnarray}
  i \frac {\partial \tilde{\Phi}} {\partial {\tilde t}} = 
  - \frac {1} {2} \frac {\partial^2 {\tilde \Phi}} {\partial {\tilde x}^2}
  + |\tilde{\Phi}|^2 {\tilde \Phi} - |{\tilde \Phi}| {\tilde \Phi},
\end{eqnarray}
where ${\tilde t} = t/t_0$ and ${\tilde x} = x/x_0$. Furthermore, we introduce the
unit of $N$, $N_0 = \Phi_0^2 x_0 = (\sqrt{\pi}/2) (g/\delta g)^{3/2}$. 

The corresponding time-independent order parameter ${\tilde \Psi}({\tilde x})$, 
where ${\tilde \Phi}({\tilde x}, {\tilde t}) = {\tilde \Psi}({\tilde x}) 
e^{-i {\tilde \mu} {\tilde t}}$ satisfies the equation
\begin{eqnarray}
 - \frac {1} {2} \frac {\partial^2 {\tilde \Psi}} {\partial {\tilde x}^2}
  + |\tilde{\Psi}|^2 {\tilde \Psi} - |{\tilde \Psi}| {\tilde \Psi} = 
  {\tilde \mu} {\tilde \Psi},
 \label{tdcc}
\end{eqnarray}
where $\tilde \mu = \mu/e_0$, with $\mu $ being the chemical potential and $e_0 = \hbar^2/
(M x_0^2)$. Regarding the value of the above parameters, considering a value $\xi = 30$ nm
\cite{rrev2}, $a_{\perp} = \xi/10 = 3$ nm, $\delta g/g = 1/10$, $M = 50$ proton masses, and 
$a_{\rm sc} = 1$ nm, then $x_0$ is on the order of tens of nanometers, $t_0$ tens of 
nanoseconds, while $N_0 \approx 30$.

From now on we work with the dimensionless quantities and drop the ``tilde" from the symbols. 
Since our goal is to investigate the yrast state, i.e., to minimize the energy fixing the 
angular momentum, we work with the following extended energy functional,
\begin{eqnarray}
 {\cal E}(\Psi, \Psi^*) = \frac {1} {2} \int \left| \frac {\partial {\Psi}} 
 {\partial {x}} \right|^2 \, dx + \frac 1 2 \int |\Psi|^4 \, dx - \frac 2 3 \int |\Psi|^3 \, dx
  \nonumber \\ 
  - \mu \int |\Psi|^2 \, dx - \Omega \int \Psi^* {\hat \ell} \Psi \, dx.
\nonumber \\
\label{tdcccc}
\end{eqnarray}
Here ${\hat \ell}$ is the operator of the angular momentum, while $\mu$ and $\Omega$ 
are Lagrange multipliers, which correspond to the two conserved quantities in our problem,
namely the (reduced) atom number $N$ and the angular momentum per particle $\ell$, 
respectively.

We minimize numerically the extended energy functional of Eq.\,(\ref{tdcccc}) with the
damped second-order-in-fictitious-time \cite{SOG}, with dynamical constraints, method 
from \cite{GO} adopted to the specific non-linear terms. Excited states are calculated 
with higher damping and hence lower accuracy, compared to yrast states (normal damping).

\section{Ground state -- Non-rotating problem}

Before we proceed to the problem of rotation, it is crucial to examine the 
non-rotating state. A trivial solution of our problem is that of a homogeneous
density distribution, $\Psi_0 = \sqrt{n_0} = \sqrt{N/(2 \pi R)}$. Depending on the 
value of the chosen parameters, the density may be homogeneous, or localized. Below 
we investigate the dynamic and energetic stability of the homogeneous solution 
under long-wavelength excitations. As we see, these two conditions lead to identical
results.

\subsection{Dynamic and energetic stability of the homogeneous solution, and speed of sound}

Starting with the dynamic stability, let us consider small deviations of the order 
parameter from the homogeneous solution, $\Psi = \Psi_0 + \delta \Psi$. The resulting 
(linearised) equation for $\delta \Psi$ is
\begin{eqnarray}
 i \frac {\partial \delta \Psi} {\partial t} =
 - \frac {1} {2} \frac {\partial^2 {\delta \Psi}} {\partial {x}^2}
 + \left(n_0 - \frac 1 2 \sqrt{n_0} \right) [\delta \Psi + (\delta \Psi)^*].
 \label{tdds}
\end{eqnarray}
Assuming plane-wave solutions, $\delta \Psi(x,t) \propto e^{i m x/R - i \omega t}$, 
we find that 
\begin{eqnarray}
  \omega^2 R^2 = m^2 \left(n_0 - \frac 1 2 \sqrt{n_0} \right) + \frac {m^4} {4 R^2},
\end{eqnarray}
where the periodicity of the problem implies that $m$ is an integer. From the above 
equation we find that the speed of sound $c$ is (for $m=1$)\cite{PS},
\begin{eqnarray}
 c = \sqrt{n_0 - \frac 1 2 \sqrt{n_0} + \frac {1} {4 R^2}}.
\label{condd}
\end{eqnarray} 
Clearly, the condition for dynamic stability of the homogeneous solution is that
$c$ has to be real, or
\begin{eqnarray}
 n_0 - \frac 1 2 \sqrt{n_0} + \frac {1} {4 R^2} \ge 0.
\label{condd33}
\end{eqnarray} 

Turning to the energetic stability of the homogeneous solution, let us consider the 
trial order parameter 
\begin{eqnarray}
  \Psi = \frac {\sqrt{N}} {\sqrt{2 \pi R}} [c_0 + 2 c_1 \cos (x/R)],
\end{eqnarray}
where $c_0$ and $c_1$ are variational parameters, with $|c_0|^2 + 2 |c_1|^2 = 1$,
due to the normalization condition. Let us also assume that we have small deviations 
from the homogeneous solution, i.e., $|c_0| \approx 1$ and $|c_1| \ll 1$. We evaluate 
the energy and expand it in the small parameter $|c_1|$. The resulting term that 
multiplies $|c_1|^2$ is $\propto 4 n_0 - 2 \sqrt{n_0} + 1/R^2$. Therefore, the 
condition for the homogeneous solution to be energetically stable coincides with 
the one of Eq.\,(\ref{condd33}), i.e., the condition for dynamic stability. 

\begin{figure}[t]
\includegraphics[width=8cm,height=6cm,angle=-0]{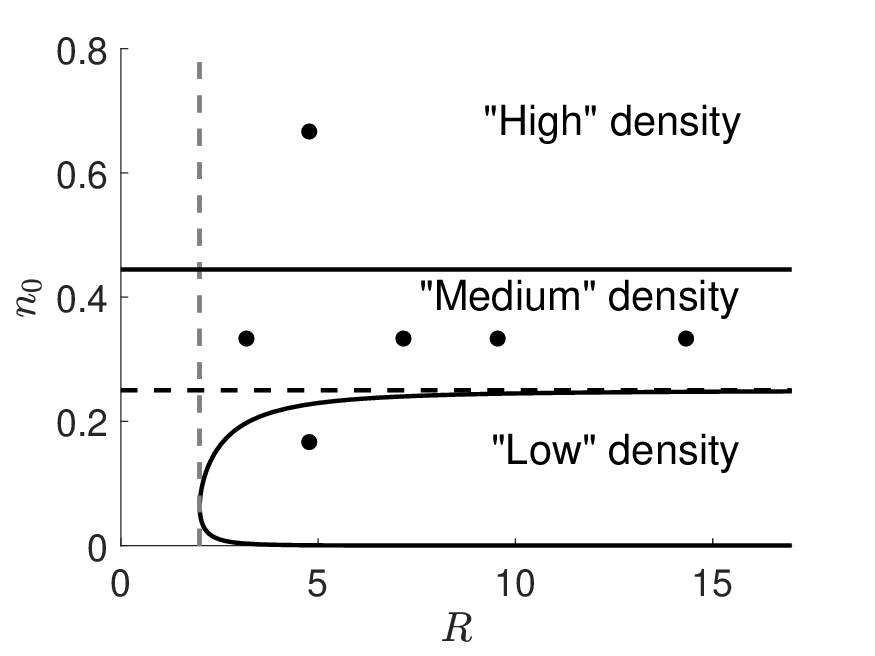}
\caption{The phase diagram for the long-wavelength energetic and dynamic stability of 
the homogeneous solution, see Eq.\,(\ref{condd33}); the lower right part of the phase
diagram is the one where the density is localized. On the horizontal axis we have the
radius of the ring $R$ and on the vertical axis $n_0 = N/L$. The (black) dots correspond 
to the cases that we have considered in Sec.\,IV, i.e., ($N = 5, L = 30$) lowest dot; 
($N = 20/3, L = 20$) middle, left dot; ($N = 15, L = 45$) middle, second dot from the 
left; ($N = 20, L = 60$) middle, third dot from the left; ($N = 30, L = 90$) middle,  
right dot; ($N = 20, L = 30$) highest dot. The vertical line denotes the value of 
$R = 2$, below which the density is always homogeneous. The horizontal dashed line 
shows the value of $n_0 = 1/4$ which separates the homogeneous from the localized 
phase for $R \to \infty$. This is also the boundary between ``low" and ``medium" 
values of $n_0$. The horizontal solid line shows the value of $n_0 = 4/9$, which 
is the boundary between ``medium" and ``high" values of $n_0$. The density is 
measured in units of $|\Phi_0|^2$, $N$ in units of $N_0$, and the length in units 
of $x_0$.}
\end{figure}

Figure 1 shows the phase diagram that we derived above, i.e., Eq.\,(\ref{condd33}),
where in the lower right part of the phase diagram the density is localized. In 
the thermodynamic limit, $N \to \infty$ and $L = 2 \pi R \to \infty$, with $n_0 = 
N/L$ finite, we get the horizontal, asymptotic, straight (dashed) line, $n_0 = 1/4$, 
which is the critical value of $n_0$ in order for the homogeneous solution to be 
stable in a ``large" ring. This value of $n_0 = 1/4$ is the one that separates the 
regime of ``low" density from the regime of ``medium" density, in our terminology; 
see below. For ``large", but finite values of $R$, the first-order correction to 
the value $n_0 = 1/4$ that follows from Eq.\,(\ref{condd33}) is $n_0 \approx 
1/4-1/(2 R^2)$. 

The other horizontal (solid) line separates the region between ``medium" and 
``high" density, see also below. Finally, it follows from Eq.\,(\ref{condd33}) 
that there is a minimum value of $R$ below which the density is homogeneous for 
all values of $n_0$, which is $R = 2$. This is shown as the vertical line in 
Fig.\,1.

\begin{figure}[h]
\includegraphics[width=8cm,height=6cm,angle=-0]{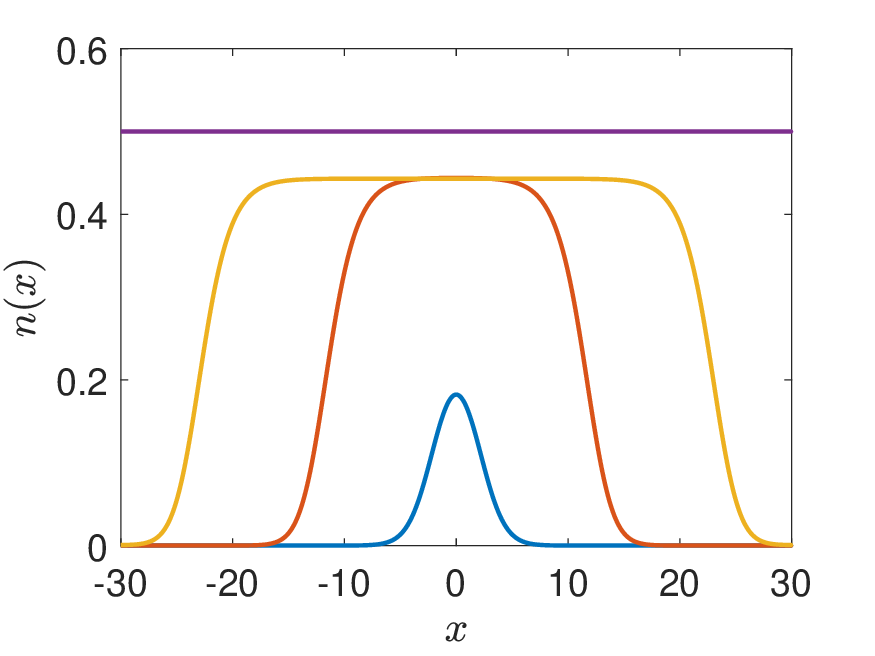}
\caption{The density $n(x) = |\Psi(x)|^2$ of the order parameter for $N = 1$ (most
narrow curve), 10 (intermediate curve), 20 (most wide curve), and 30 (homogeneous), 
with $L = 2 \pi R = 60$ and $\ell = 0$. The density is measured in units of $|\Phi_0|^2$,
$N$ in units of $N_0$, and the length in units of $x_0$.}
\end{figure}

\subsection{Transition from ``small" to ``large" droplets -- Thomas-Fermi limit}

For zero angular momentum and small $N$ the density distribution has a Gaussian shape 
and as $N$ increases, the droplet becomes wider \cite{th1}. Eventually, when the size 
of the droplet approaches the size of the circumference of the ring, the density becomes 
homogeneous. An example of this is shown in Fig.\,2, where we have plotted the density 
for $N = 1, 10, 20$, and 30, with $L = 2 \pi R = 60$. 

For $N \gg 1$ the droplet may develop a ``flat-top" density distribution (see, e.g., 
the case $N = 20$ in Fig.\,2), provided that its size is relatively smaller than 
$L = 2 \pi R$. Then, the kinetic energy is negligible and the Thomas-Fermi approximation 
is valid, which implies that the total energy per particle is
\begin{eqnarray}
 \frac {E_{TF}} N = \frac 1 2 n_{d} - \frac 2 3 \sqrt{n_{d}},
\label{tfapp}
\end{eqnarray}
where $n_d = N/d$ is the density of the flat-top state. Minimization of the above 
expression with respect to $n_{d}$ implies that $n_{d} = 4/9$. The minimized energy 
per particle, which is also equal to the chemical potential $\mu$, is $E_{TF}/N = 
\mu_{TF} = - 2/9$. Finally, the size of the droplet is $N/n_d = 9N/4$ in this limit
and therefore the condition $d \ll L = 2 \pi R$ takes the form $9N/4 \ll L = 2 \pi R$. 
This condition may also be written as $n_0 \ll 4/9$, with $N \gg 1$.

In the previous subsection we derived the condition $n_0 < 1/4$ of ``low" density. 
From the above analysis it follows that the value of $n_0$ which separates the 
regimes between ``medium" and ``high" values is $n_0 = 4/9$. When $n_0 > 4/9$, 
we enter the regime of ``high" density.  Thus, the ``medium" range of the density 
is $1/4 < n_0 < 4/9$, which is particularly interesting, as we explain below 
\cite{POL}. The three regimes of ``low", ``medium", and ``high" values of $n_0$
are shown in the phase diagram of Fig.\,1.  

We should stress that for medium values of the density, $1/4 < n_0 < 4/9$ the 
energy per particle of Eq.\,(\ref{tfapp}) is a decreasing function of $n_0$ 
[see Eq.\,(\ref{tfapp})], reaching its minimum for $n_0 = 4/9$. Another 
peculiar feature of the system in this range of $n_0$ is that the pressure 
$P = - dE/dV = n^2 d(E/N)/dn$, is negative, since
\begin{equation}
  P = n_d^2 \left( \frac 1 2 - \frac 1 {3 \sqrt{n_{d}}} \right).
\end{equation}

\section{Rotating problem}

Now that we have a clear picture of the lowest-energy state(s) of a droplet in the 
absence of any rotation, we can proceed to the problem of a nonzero value of the 
angular momentum. As we mentioned also above, we are interested in the lowest-energy 
state of a quantum droplet, for some fixed value of the (expectation value of) the 
angular momentum, $E(\ell)/N$. We should stress at this point that if one wants to 
work with a fixed rotational velocity of the ring $\Omega$, this may be done easily 
by examining the minimum of $E(L)/N$ in the rotating frame, i.e., 
$E_{\rm rot}/N = E(\ell)/N - \ell \Omega$.

We will focus on the three regimes which were described in the end of the previous
section, which give qualitatively different results. The first one is that of 
droplets of ``low" density, where $n_0 = N/(2 \pi R) < 1/4$. The other limiting 
case is that of droplets of ``high" density, where $n_0 > 4/9$. Finally, we examine 
a third regime, where $1/4 < n_0 < 4/9$, which is the most interesting one. 

The (black) dots in Fig.\,1 show the cases that we have considered, plotted on the 
corresponding phase diagram which shows the phase boundary for long-wavelength 
energetic and dynamic stability of the homogeneous solution. 

\subsection{Droplets of ``low" density}

As we have seen, in the limit of a ``low" value of $n_0$, $n_0 < 1/4$, the density 
of a droplet is always localized and as a result its rotational response resembles 
that of a scalar condensate with an attractive effective interaction. 

In this limit, the droplet carries the angular momentum via center-of-mass excitation, 
for all values of $\ell$. As a result, the density distribution is not affected by 
the angular momentum, as seen in Fig.\,3. The corresponding energy per particle, 
i.e., the dispersion relation $E(\ell)/N$ (bottom plot in Fig.\,3), has the following,
exact, form due to the excitation of the center of mass,
\begin{equation}
   \frac {E(\ell)} N = \frac {E(\ell=0)} N + \frac {\ell^2} {2 R^2}.
   \label{parabola}
\end{equation}
Therefore, in this limit the motion is essentially classical.

Figure 3 shows the density and the phase of the order parameter, for $N = 5$ and 
$L = 2 \pi R = 30$ (see the lowest dot in the phase diagram of Fig.\,1)
and three values of $\ell$. Clearly the density is the same, while it is only the
phase that changes. For these parameters $n_0 = 1/6 < 1/4$. For this low value 
of $n_0$ only the localized solution is present for $\ell = 0$, i.e., the 
homogeneous solution does not exist, not even as a metastable state.
 
\begin{figure}[t]
\includegraphics[width=5.5cm,height=3.7cm,angle=-0]{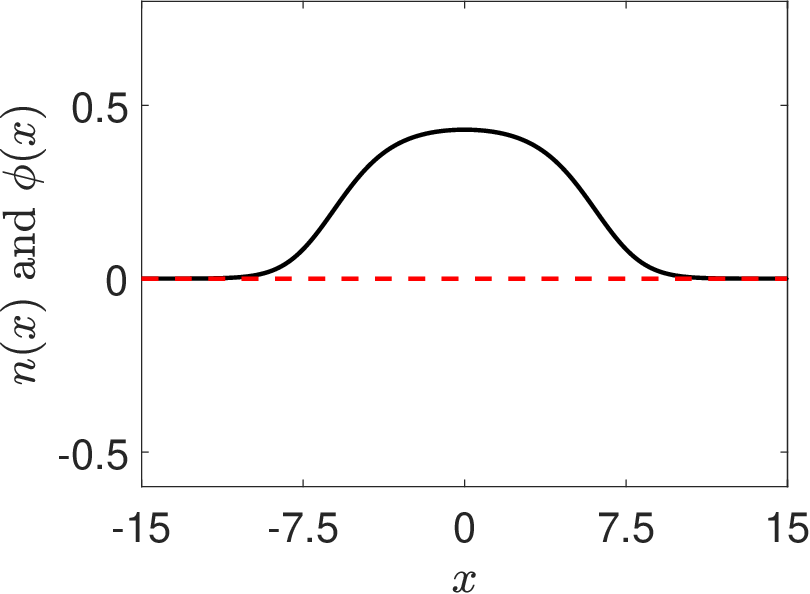}
\includegraphics[width=5.5cm,height=3.7cm,angle=-0]{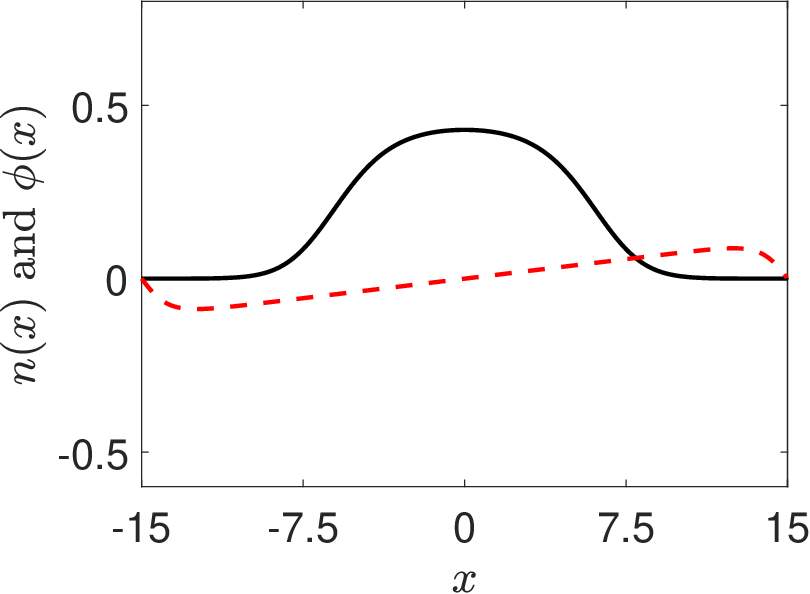}
\includegraphics[width=5.5cm,height=3.7cm,angle=-0]{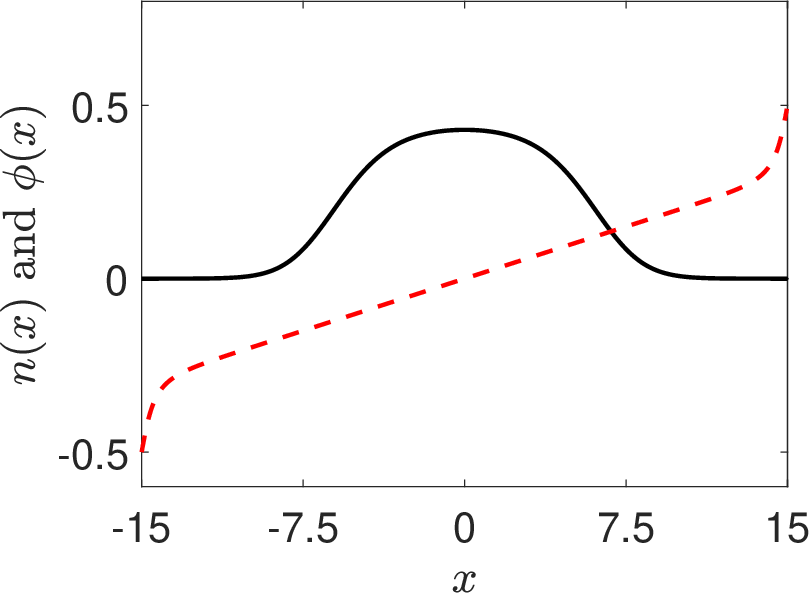}
\includegraphics[width=5.5cm,height=3.7cm,angle=-0]{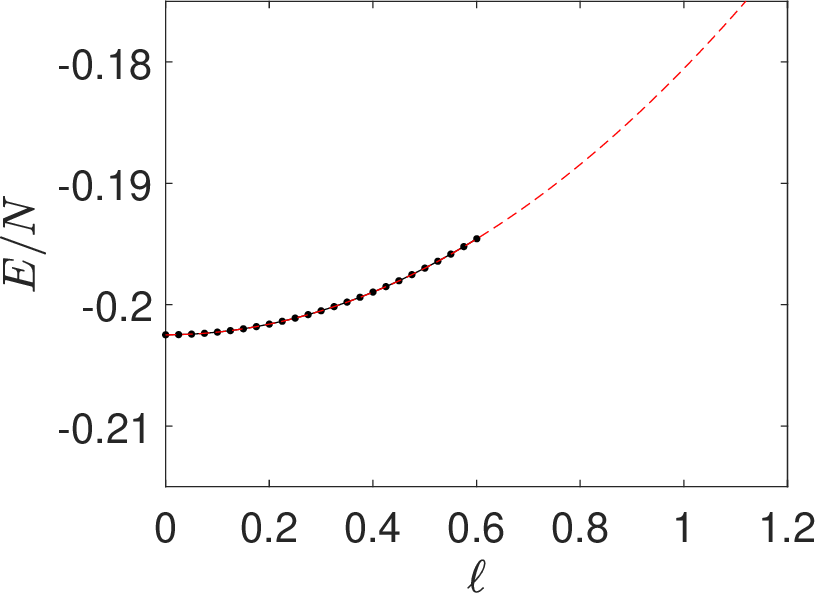}
\caption{Data for ``low" droplet density, with $N = 5$ and $R = 15/\pi$, 
i.e., $n_0 = 1/6$; lowest dot in the phase diagram of Fig.\,1. The density 
$n(x)$ (solid curve) and the phase $\phi(x)$ divided by $2 \pi$ (dashed curve) 
of the order parameter $\Psi(x) = \sqrt{n(x)} e^{i \phi(x)}$, for (from top 
to bottom) $\ell = 0, 0.2250$, and 0.6004. The plot at the bottom shows the 
corresponding dispersion relation, which is fitted by the function $E(\ell)/N 
= E(\ell=0)/N + \ell^2/(2R^2)$, as expected (red, dashed line). The density 
is measured in units of $|\Phi_0|^2$, $N$ in units of $N_0$, the length in 
units of $x_0$, the energy is units of $e_0$ and the angular momentum in units 
of $\hbar$.}
\end{figure}

\begin{figure}[t]
\includegraphics[width=5.5cm,height=3.3cm,angle=-0]{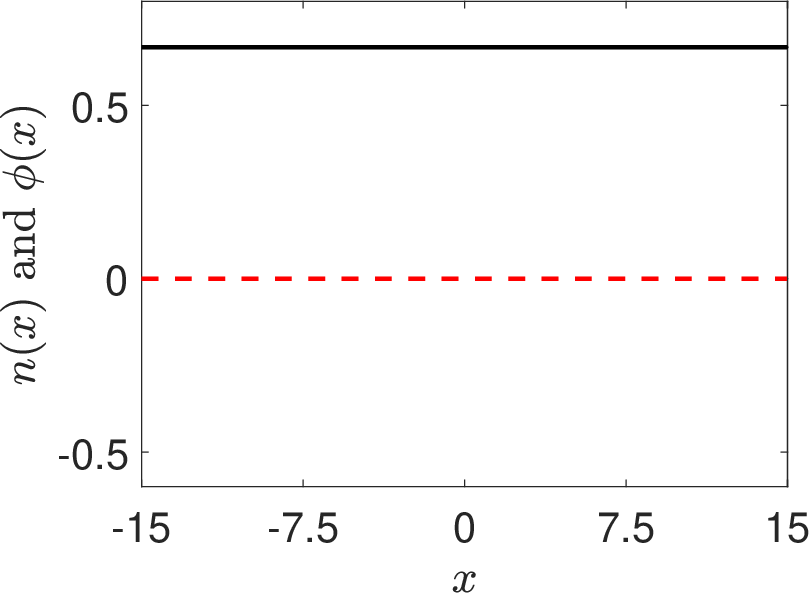}
\includegraphics[width=5.5cm,height=3.3cm,angle=-0]{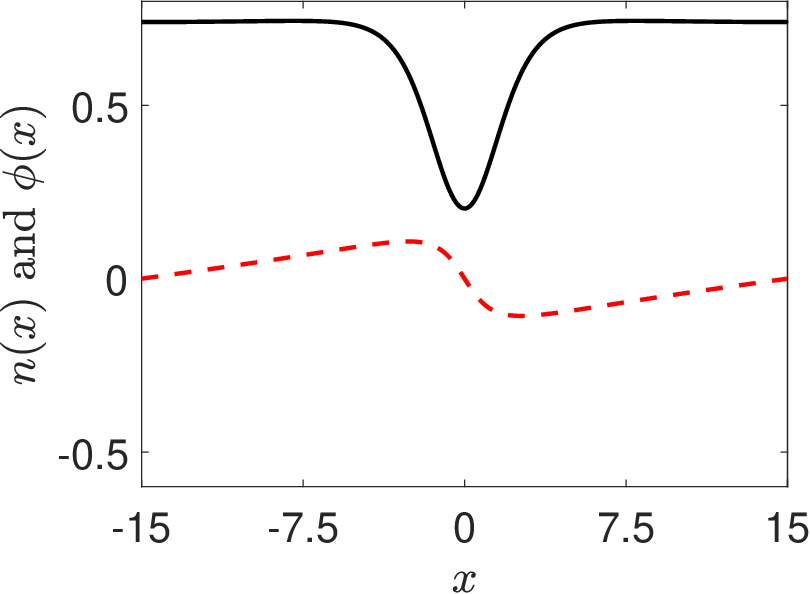}
\includegraphics[width=5.5cm,height=3.3cm,angle=-0]{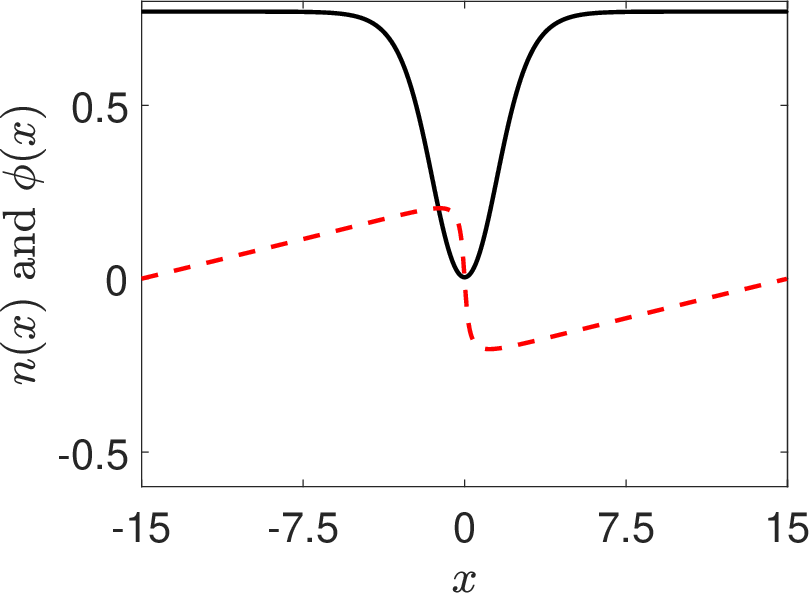}
\includegraphics[width=5.5cm,height=3.3cm,angle=-0]{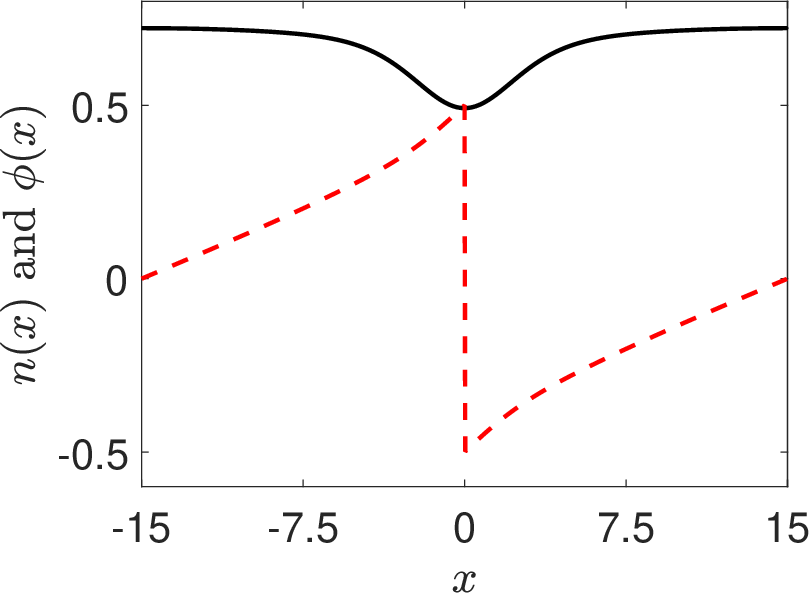}
\includegraphics[width=5.5cm,height=3.7cm,angle=-0]{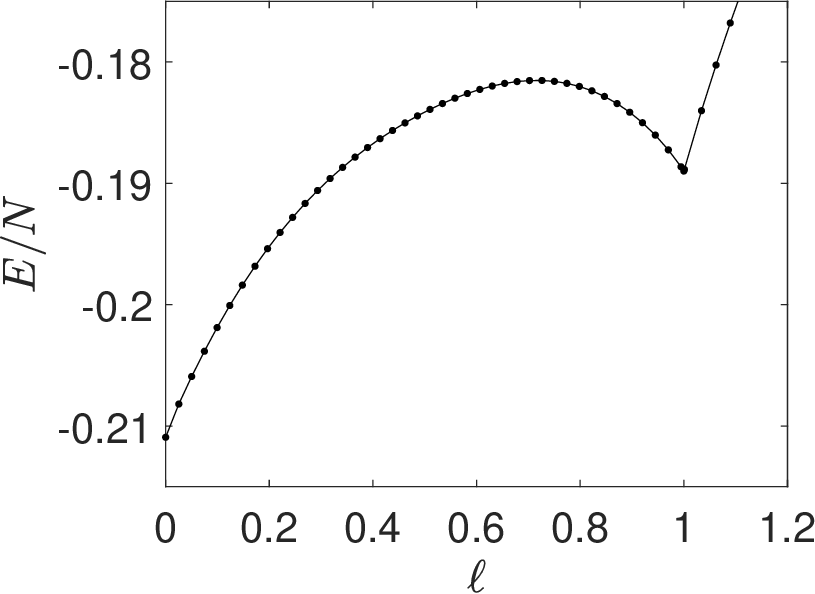}
\caption{Data for ``high" droplet density, with $N = 20$ and $R = 15/\pi$, 
i.e., $n_0 = 2/3$; highest dot in the phase diagram of Fig.\,1. The density 
$n(x)$ (solid curve) and the phase $\phi(x)$ divided by $2 \pi$ (dashed curve) 
of the order parameter $\Psi(x) = \sqrt{n(x)} e^{i \phi(x)}$, for (from top 
to bottom) $\ell = 0, 0.1482, 0.4860$, and 0.9700. The plot at the bottom 
shows the corresponding dispersion relation. The density is measured in units 
of $|\Phi_0|^2$, $N$ in units of $N_0$, the length in units of $x_0$, the 
energy is units of $e_0$ and the angular momentum in units of $\hbar$.}
\end{figure}

\subsection{Droplets of ``high" density}

Let us now turn to the opposite limiting case of a droplet with a ``high" density
$n_0$. In this case for zero angular momentum the density is homogeneous and the 
problem resembles that of a scalar condensate, with a repulsive interatomic interaction. 
Also, only the homogeneous solution is present for $\ell = 0$, i.e., the localized 
solution is not present, not even as a metastable state. 

For small values of the angular momentum we have plane-wave-like solutions, where
there is a small variation from the homogeneous density distribution. The dispersion 
relation is linear in $\ell$, with a slope which is equal to the speed of sound $c$,
given by Eq.\,(\ref{condd}). 

As $\ell$ increases, the minimum of the density decreases, up to $\ell = 1/2$, where 
we have a node in the density. In the language of solitary-wave excitation, this is a 
``dark" solitary wave. Due to the finiteness of the ring potential for $\ell = 1/2$ 
the propagation velocity of the solitary wave is nonzero (which is the slope of the
dispersion relation at this value of $\ell$) \cite{SMKJ}. For even higher values of 
$\ell$, the minimum of the density increases, due to Bloch's theorem \cite{FB}, which 
implies that the density is the same for any two solutions with angular momentum 
$\ell$ and $1 - \ell$, with $0 \le \ell \le 1$. For some value of $\ell$ between 1/2 
and 1 the slope of the dispersion relation vanishes and therefore the wave is static 
(but not ``dark"), as in a scalar condensate (with an effective repulsive interaction). 
Finally, for $\ell = 1$ the density is again homogeneous (as for $\ell = 0$), and the 
phase is a linear function of $\ell$, having a difference of $2 \pi$. For $\ell > 1$ 
the problem is essentially the same due to Bloch's theorem \cite{FB}: The solution is 
the one for $\ell < 1$, via excitation of the center-of-mass. 

We stress that in the thermodynamic limit $N \to \infty$, $L = 2 \pi R \to \infty$, 
with $N/L$ finite, the dark wave becomes also static \cite{SMKJ}. At this value 
of $\ell = 1/2$ the phase also develops a jump by $2 \pi$. 

Regarding the dispersion relation, this has a negative curvature and is a periodic 
function, on top of a parabola, as in scalar condensates with an repulsive effective
interaction (according to Bloch's theorem \cite{FB}). 

A few examples are shown in Fig.\,4, for $N = 20$ and $L = 2 \pi R = 30$ (see the 
highest dot in the phase diagram of Fig.\,1). For these parameters $n_0 = 2/3 
> 4/9$. 

Finally, let us make a remark about the stability of persistent currents. From the 
dispersion relation of Fig.\,4 it is clear that there is a metastable minimum at $\ell 
= 1$. This implies that the system supports persistent currents, as in repulsive scalar 
condensates. For $n_0 > 4/9$ (where we have the ``traditional" dispersion relation, 
with a negative curvature), the slope of the dispersion relation $c'$ for $\ell \to 
1^-$ is 
\begin{eqnarray}
 c' = \frac 1 R - \sqrt{n_0 - \frac 1 2 \sqrt{n_0} + \frac {1} {4 R^2}}.
\end{eqnarray} 
For the stability of persistent currents one has to solve the equation
\begin{equation}
  n_0 - \frac 1 2 {\sqrt{n_0}} - \frac 3 {4 R^2} = 0. 
\end{equation}
For ``large" $R$ we get the approximate result
\begin{eqnarray}
 n_0 \approx \frac 1 4 + \frac {3} {2 R^2}.
\end{eqnarray} 

\subsection{Droplets of ``medium" density} 

Finally, we turn to the third case that we consider, namely the regime of 
``medium" density, which is actually the most interesting one, for the reasons 
that follow below. In this case we move horizontally in the phase diagram of 
Fig.\,1, fixing $n_0$ to the value of 1/3 and we choose four different values 
of $R$. 

One general observation that we have here is that the energy of the state 
with a homogeneous density distribution depends on $n_0$ only and as a result
in all the four cases that we consider $E(\ell = 0)/N$, which we denote as 
$E_{\rm sol}(\ell = 0)/N$, is the same and equal to $(1/2) n_0-(2/3) \sqrt{n_0} 
\approx -0.2182$ [see Eq.\,(\ref{tfapp})]. Excitations above the homogeneous 
state for small values of $\ell$ are plane-wave solutions, with a linear 
dispersion relation, and a corresponding speed of sound $c$ which is given 
by Eq.\,(\ref{condd}). Clearly, as $R$ increases, $c$ decreases, or in other 
words the slope of the dispersion relation at $\ell \to 0^+$ decreases.

In addition to the solution with a homogeneous density distribution we have 
the state with a localized density distribution (we refer to the case with 
$\ell = 0$). The energy of this state at $\ell = 0$, which we denote as 
$E_{\rm COM}(\ell = 0)/N$, does depend on $R$ and is also a decreasing 
function of $R$. For the three smallest values of $R$ that we consider, 
$E_{\rm sol}(\ell = 0) < E_{\rm COM}(\ell = 0)$, while for the largest 
value $E_{\rm sol}(\ell = 0) > E_{\rm COM}(\ell = 0)$. Finally, we stress
that, contrary to the case of a homogeneous density distribution, here 
we have center-of-mass excitation, with a parabolic dispersion relation.

In the first case that we choose $R = 10/\pi$, and $N = 20/3$ (middle, 
left dot in the phase digram of Fig.\,1). Here the yrast state for $\ell 
= 0$ is the homogeneous solution. Furthermore, the energy scale $1/(2R^2)$, 
i.e., the kinetic energy, is comparable with the energy of the nonlinear 
terms. As a result the dispersion relation, shown in the bottom plot of 
Fig.\,5, is close to a linear function, while the density is well 
approximated by a sinusoidal function (top three plots in Fig.\,5). In 
addition, we have the center-of-mass states, which have a higher energy, 
though; see the fourth and the fifth plots from the top in Fig.\,5. 
As seen from the dispersion relation (bottom plot of Fig.\,5), their 
energy, which is denoted as rings, is higher. 

\begin{figure}[t]
\includegraphics[width=5.cm,height=2.7cm,angle=-0]{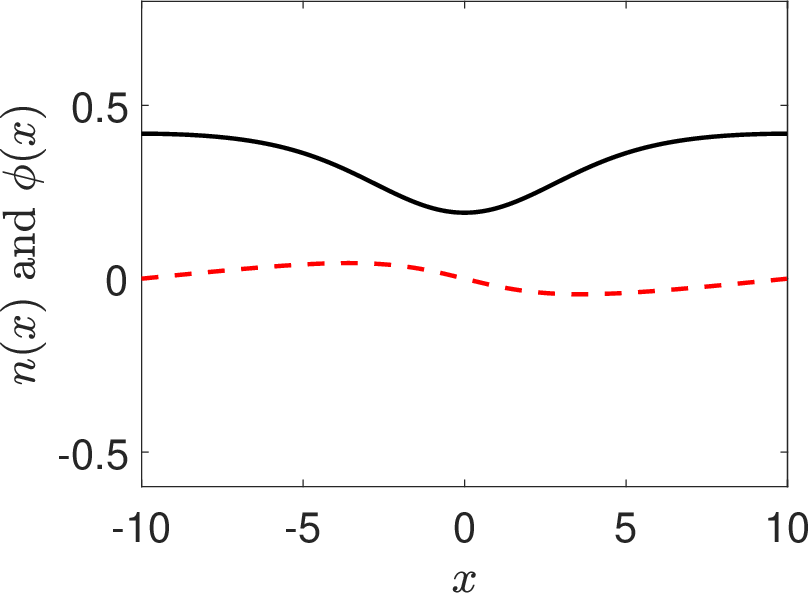}
\includegraphics[width=5.cm,height=2.7cm,angle=-0]{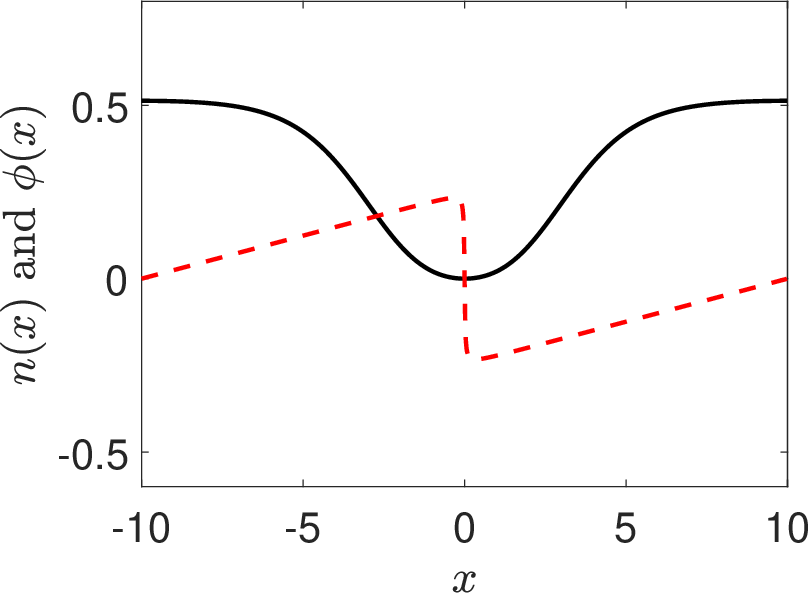}
\includegraphics[width=5.cm,height=2.7cm,angle=-0]{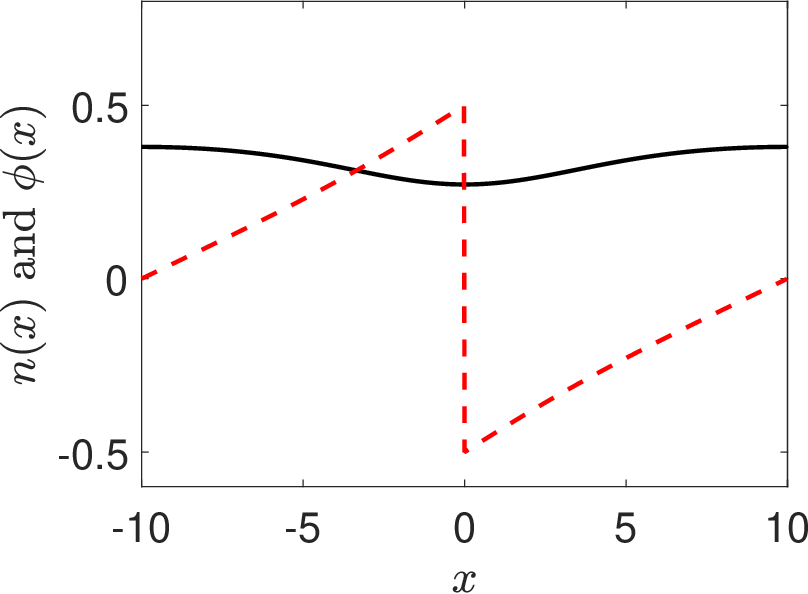}
\includegraphics[width=5.cm,height=2.7cm,angle=-0]{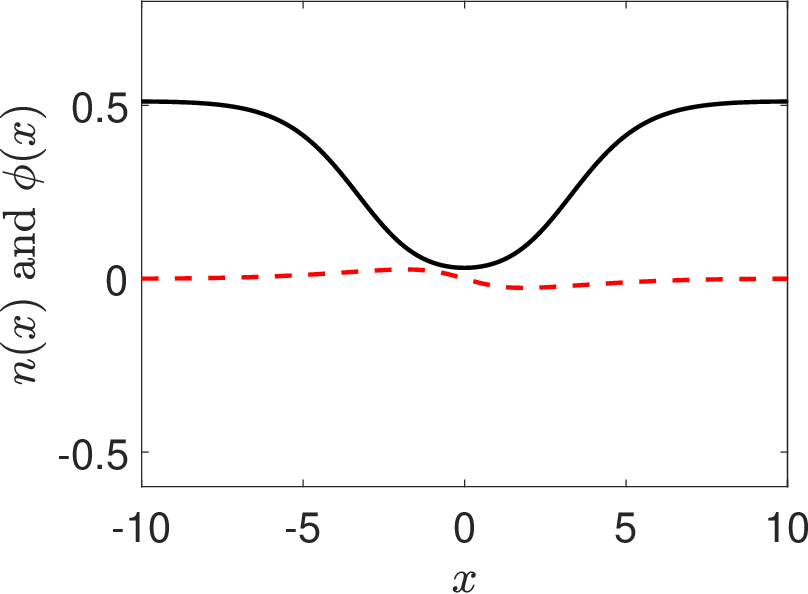}
\includegraphics[width=5.cm,height=2.7cm,angle=-0]{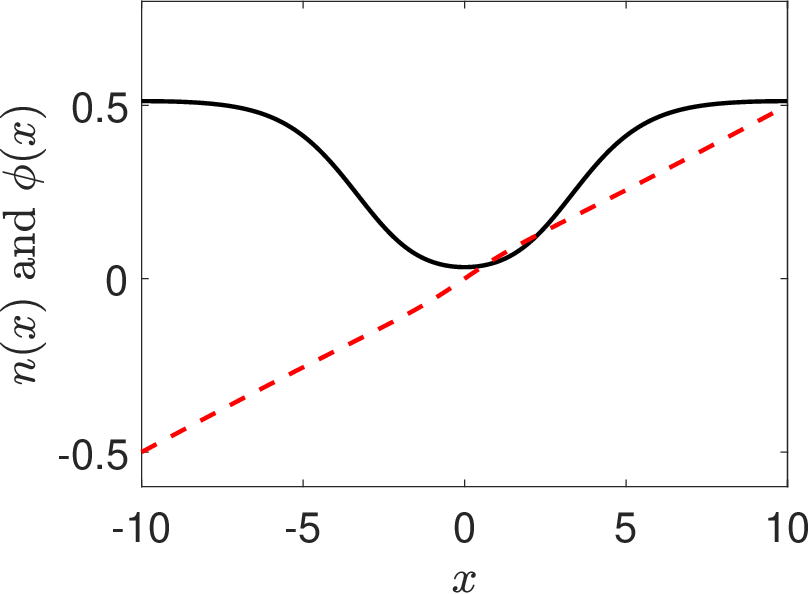}
\includegraphics[width=5.cm,height=2.7cm,angle=-0]{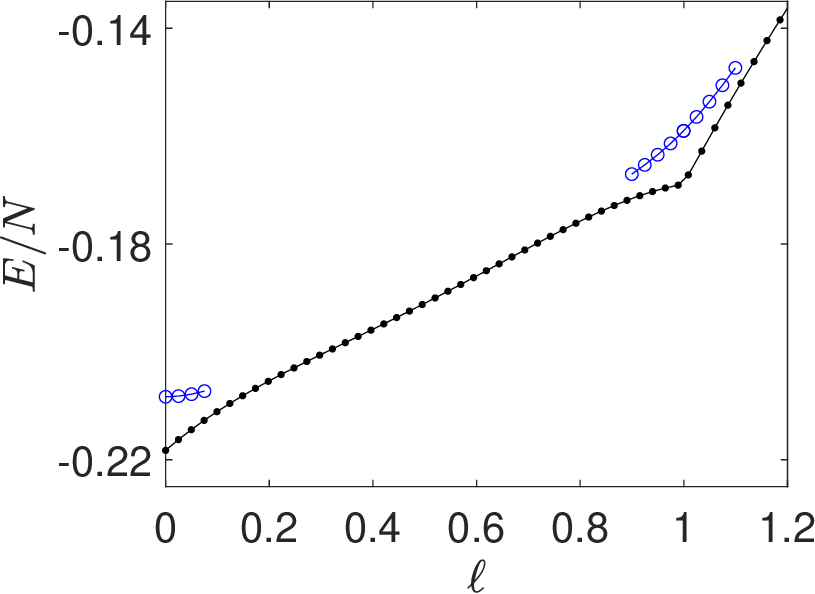}
\caption{Data for ``medium" droplet density, with $N = 20/3$ and $R = 10/\pi$,
i.e., $n_0 = 1/3$; middle, left dot in the phase diagram of Fig.\,1. The 
density $n(x)$ (solid curve) and the phase $\phi(x)$ divided by $2 \pi$ 
(dashed curve) of the order parameter $\Psi(x) = \sqrt{n(x)} e^{i \phi(x)}$, 
for (from top to bottom) ($\ell = 0.0496, 0.4952$, and 0.9890) for the 
yrast state, and ($\ell = 0.0500$, and 0.9745) for the excited 
(center-of-mass excitation) states, denoted as rings at the bottom plot, 
which shows the dispersion relation. The density is measured in units of 
$|\Phi_0|^2$, $N$ in units of $N_0$, the length in units of $x_0$, the 
energy is measured in units of $e_0$, and the angular momentum in units 
of $\hbar$.}
\end{figure} 

In the second case, shown in Fig.\,6, where $R = 45/(2 \pi)$ and $N = 15$ (middle, 
second dot from the left, in the phase diagram of Fig.\,1), the kinetic energy is 
substantially smaller than the energy of the nonlinear terms. As a result, the 
density distribution is no longer approximated by a sinusoidal function. The 
yrast states for $\ell=0$ and $\ell=1$ are the usual plane wave states, $\Psi
= {\sqrt N}/\sqrt{2 \pi R}$ and $\Psi = {\sqrt N} e^{i x/R}/\sqrt{2 \pi R}$, 
respectively, with a homogeneous density distribution. The dispersion relation 
for small values of $\ell$ has the expected linear behavior, with a slope which 
is equal to the speed of sound $c$ (sound-wave excitation). A similar picture 
also develops for $\ell$ smaller, but close to unity, due to Bloch's theorem 
\cite{FB}. 

The most interesting observation here is associated with the curvature of the 
dispersion relation. For small values of $\ell$ the curvature is negative. 
However, as $\ell$ increases, the curvature becomes positive, and the spectrum 
resembles the phonon-roton spectrum; see, e.g. Ref.\,\cite{PS}. In this part of 
the spectrum with positive curvature the motion also resembles center-of-mass-like 
excitation; see the two highest plots in Fig.\,6. The third and the fourth plots 
from the top in Fig. 6 show examples of the (excited) center-of-mass states for 
$\ell=0$, and $\ell=1$, respectively. In the dispersion relation shown at the 
bottom of Fig. 6 the latter are among the (blue) rings.  

\begin{figure}[h]
\includegraphics[width=5cm,height=3.cm,angle=-0]{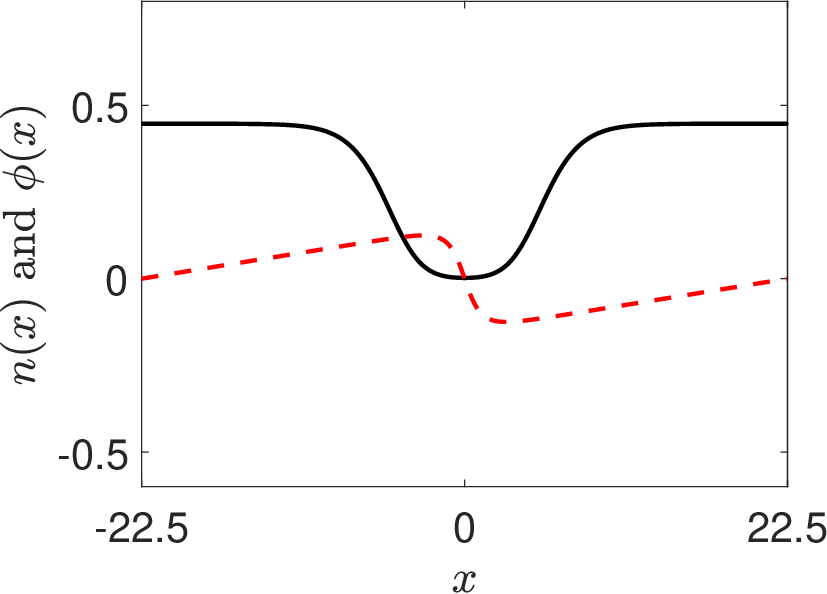}
\includegraphics[width=5cm,height=3.cm,angle=-0]{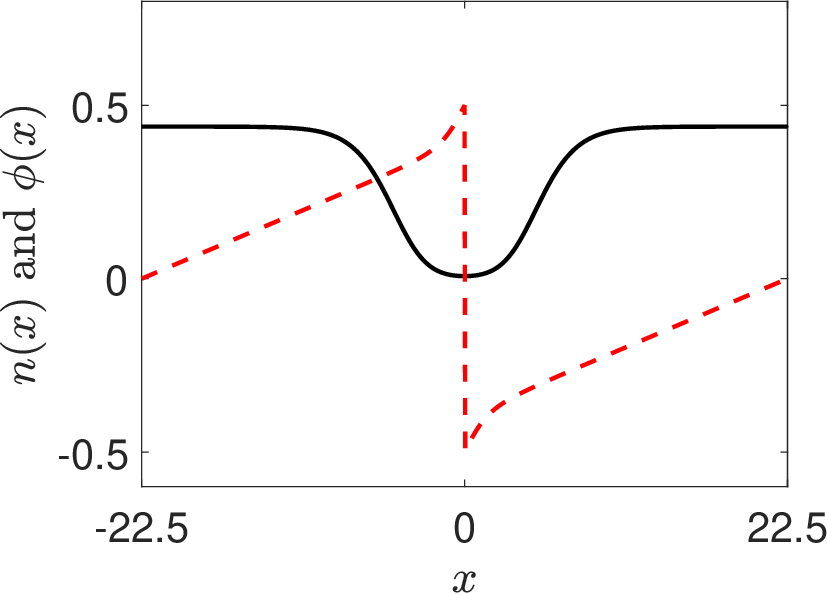}
\includegraphics[width=5cm,height=3.cm,angle=-0]{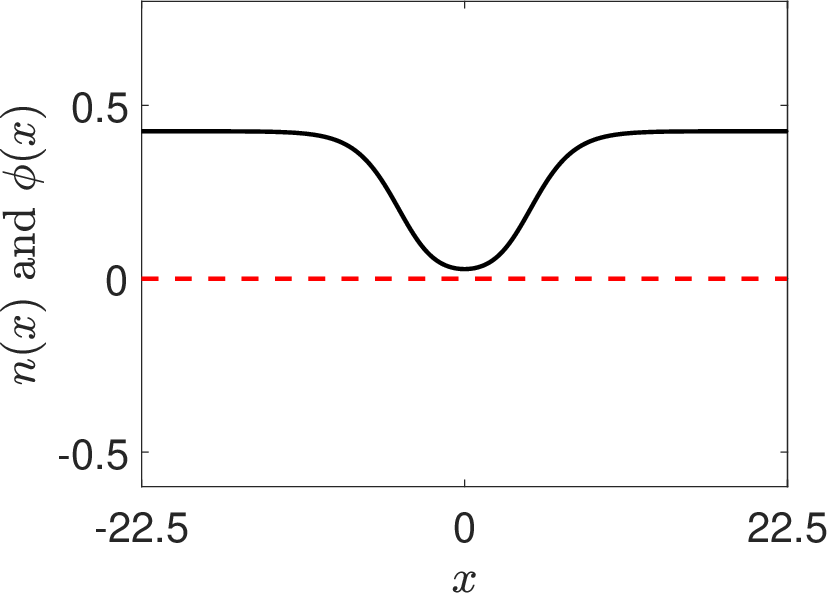}
\includegraphics[width=5cm,height=3.cm,angle=-0]{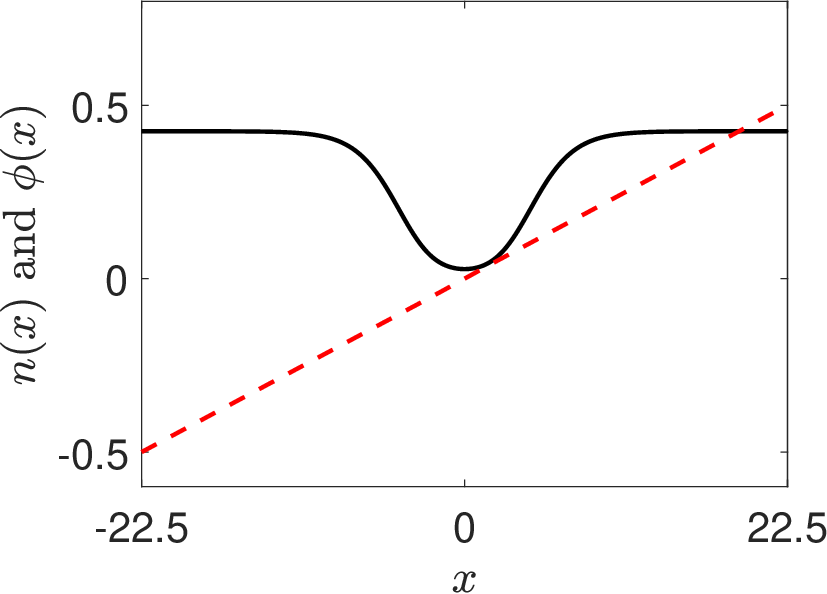}
\includegraphics[width=5.5cm,height=3.7cm,angle=-0]{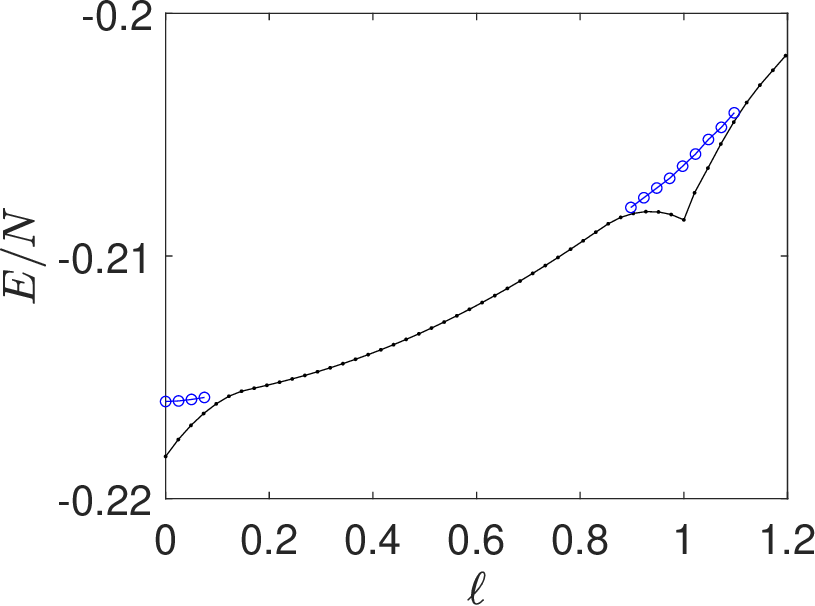}
\caption{Data for ``medium" droplet density, with $N = 15$ and $R = 45/(2 \pi)$,
i.e., $n_0 = 1/3$; middle, second dot from the left in the phase diagram of 
Fig.\,1. The density $n(x)$ (solid curve) and the phase $\phi(x)$ divided by 
$2 \pi$ (dashed curve) of the order parameter $\Psi(x) = \sqrt{n(x)} 
e^{i \phi(x)}$, for (from top to bottom) ($\ell = 0.2931$ and 0.8058) for 
the yrast state, and ($\ell = 0$, and 0.9976) for the excited (center-of-mass 
excitation) states, denoted as rings at the bottom plot, which shows the 
dispersion relation. The density is measured in units of $|\Phi_0|^2$, $N$ 
in units of $N_0$, the length in units of $x_0$, the energy is measured in 
units of $e_0$, and the angular momentum in units of $\hbar$.}
\end{figure} 

\begin{figure}[h]
\includegraphics[width=5cm,height=3.cm,angle=-0]{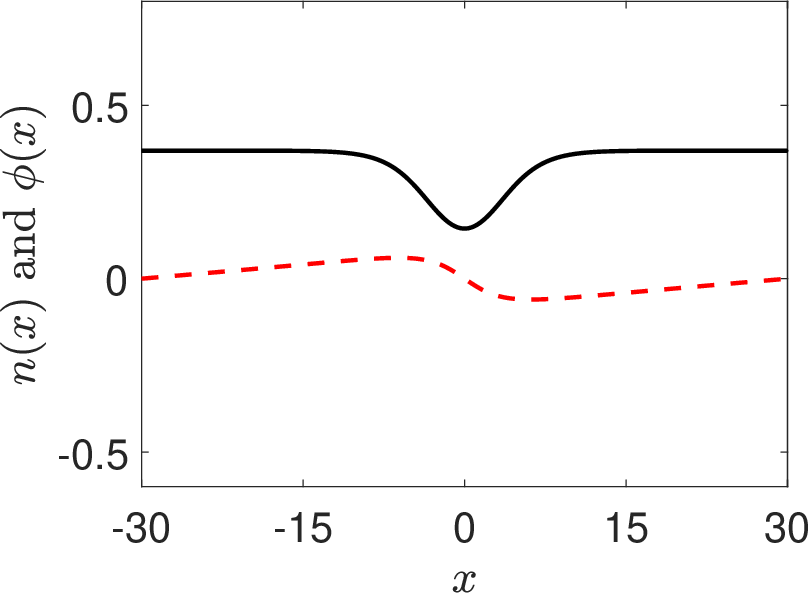}
\includegraphics[width=5cm,height=3.cm,angle=-0]{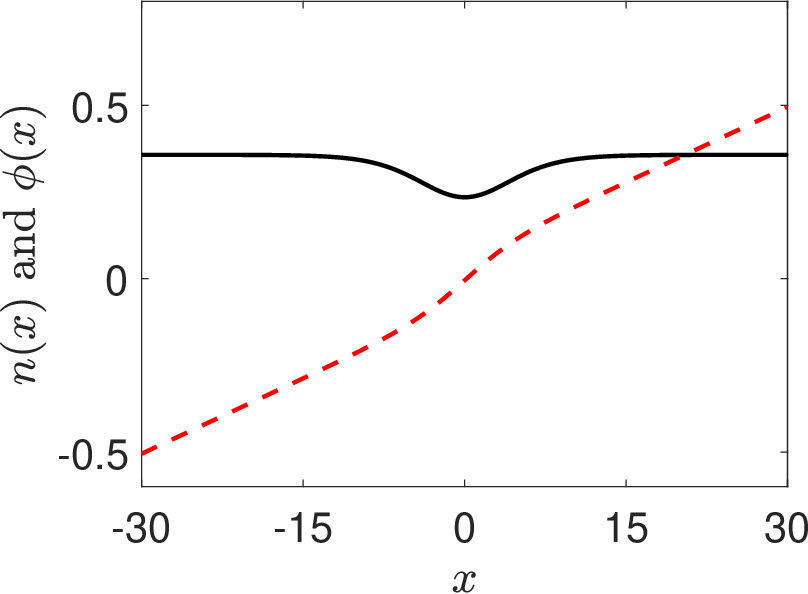}
\includegraphics[width=5cm,height=3.cm,angle=-0]{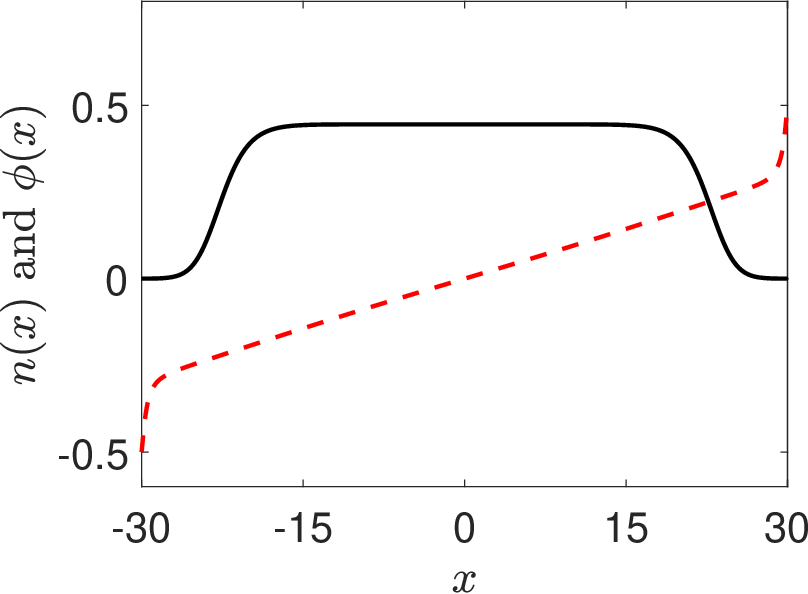}
\includegraphics[width=5cm,height=3.cm,angle=-0]{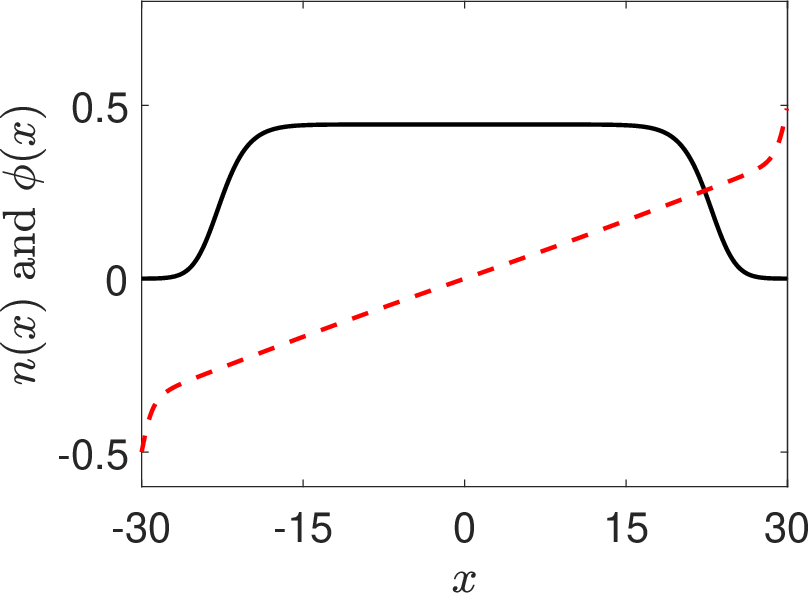}
\includegraphics[width=5.5cm,height=3.7cm,angle=-0]{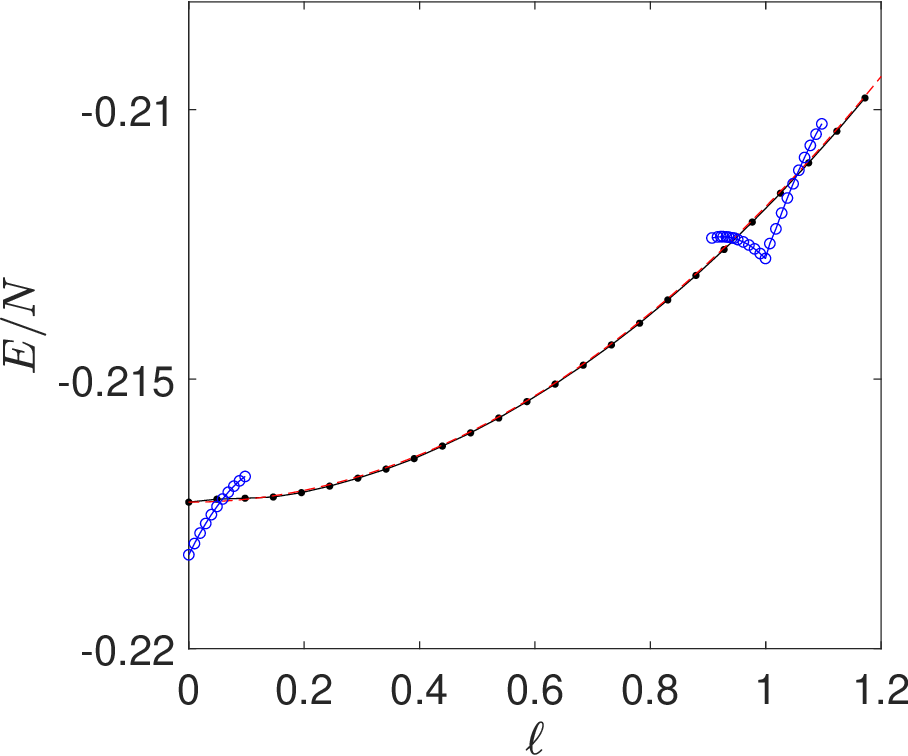}
\caption{Data for ``medium" droplet density, with $N = 20$ and $R = 30/\pi$, 
i.e., $n_0 = 1/3$; middle, third dot from the left in the phase diagram 
of Fig.\,1. The density $n(x)$ (solid curve) and the phase $\phi(x)$ 
divided by $2 \pi$ (dashed curve) of the order parameter $\Psi(x) = 
\sqrt{n(x)} e^{i \phi(x)}$, for (from top to bottom) ($\ell = 0.0685$ 
and 0.9750; see rings in the yrast curve), and ($\ell = 0.5861$ and 
0.6837; middle part of the yrast curve). The bottom plot shows the 
corresponding dispersion relation. The density is measured in units of 
$|\Phi_0|^2$, $N$ in units of $N_0$, the length in units of $x_0$, the 
energy is measured in units of $e_0$, and the angular momentum in units 
of $\hbar$.}
\end{figure}

The third case that we consider, with $R = 30/\pi$ and $N = 20$ (middle, third 
dot from the left, in the phase diagram of Fig.\,1), is also interesting. 
First of all, for zero angular momentum, the solution with a homogeneous density
is still the one with the lowest energy. As we saw above, giving angular momentum 
to the state with a homogeneous density distribution is energetically more 
expensive than that with the localized density distribution. As a result, the 
dispersion relation for the homogeneous solution increases more rapidly with 
$\ell$ than the localized one. An immediate consequence of this observation is 
that -- because for $\ell = 0$ the homogeneous solution has a lower energy than 
the localized -- there may be a level crossing for some nonzero value of the 
angular momentum of the droplet. 

Indeed, such an example is shown in Fig.\,7 (see the middle right dot in the phase 
diagram of Fig.\,1, for $n_0 = 1/3$). The two plots at the top show the density and 
the order parameter for $\ell = 0.0685$ and 0.9806 [(blue) rings in the bottom plot 
of the dispersion relation]. The third and the fourth plots refer to $\ell = 0.5861$ 
and $0.6837$ and are solutions which correspond to center-of-mass excitation [middle 
part of the dispersion relation]. These localized solutions are well approximated by 
the Thomas-Fermi solution that was described in Sec.\,III B. The energy per atom that 
we find numerically is $-0.2173$, while $E_{TF}/N = - 2/9 \approx -0.2222$. With 
increasing $\ell$ the density distribution of the droplet does not change, since, as 
we saw in Sec.\,IV A, we have center-of-mass excitation. Also, the corresponding 
dispersion relation is parabolic, see Eq.\,(\ref{parabola}).

The bottom plot of Fig.\,7 shows the dispersion relation of the two solutions. 
It is seen clearly in this plot that if one follows the yrast line, there will 
be discontinuous phase transitions close to $\ell = 0$ and $\ell = 1$. 

For the fourth and highest value of $R$ that we consider in Fig.\,8,  
$R = 45/\pi$ and $N = 30$ (middle, right dot in the phase diagram of Fig.\,1),  
$E_{\rm sol}(\ell = 0) > E_{\rm COM}(\ell = 0)$, i.e., the homogeneous state has 
a higher energy than the localized (for $\ell = 0$, but also for all $\ell \neq 
0$). We have confirmed numerically that the energy of the homogeneous solution for 
$\ell = 0$, shown as a square point at the dispersion relation (bottom plot of 
Fig.\,8), is given by the analytic result $(1/2) n_0 - (2/3) \sqrt{n_0}$, as 
expected. Also, the other square point at $\ell = 1$ has a higher energy by 
$1/(2 R^2)$, i.e., its energy is $(1/2) n_0 - (2/3) \sqrt{n_0} + 1/(2 R^2)$, 
as Bloch's theorem implies. Therefore, in the yrast state we have pure center-of-mass
excitation of the localized droplet, while the dispersion relation is parabolic, 
as seen clearly in Fig.\,8.  

\begin{figure}[t]
\includegraphics[width=5.cm,height=3.cm,angle=-0]{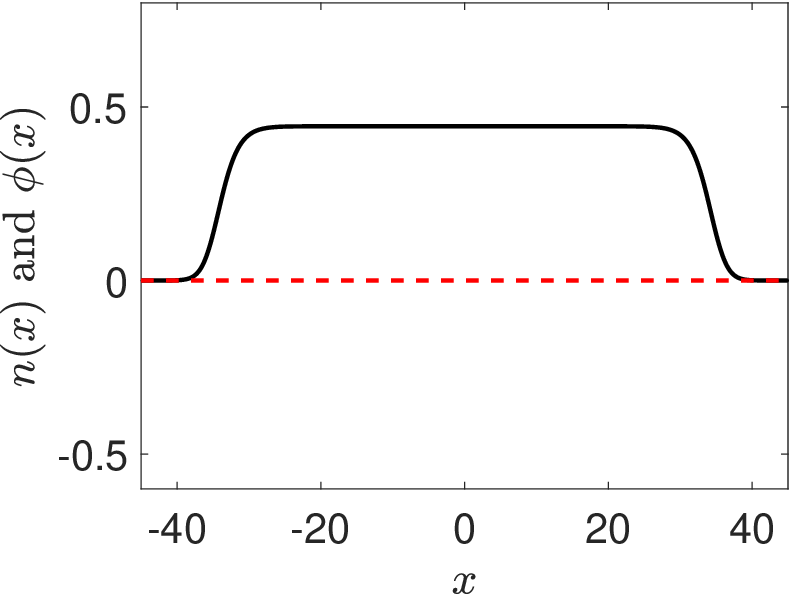}
\includegraphics[width=5.cm,height=3.cm,angle=-0]{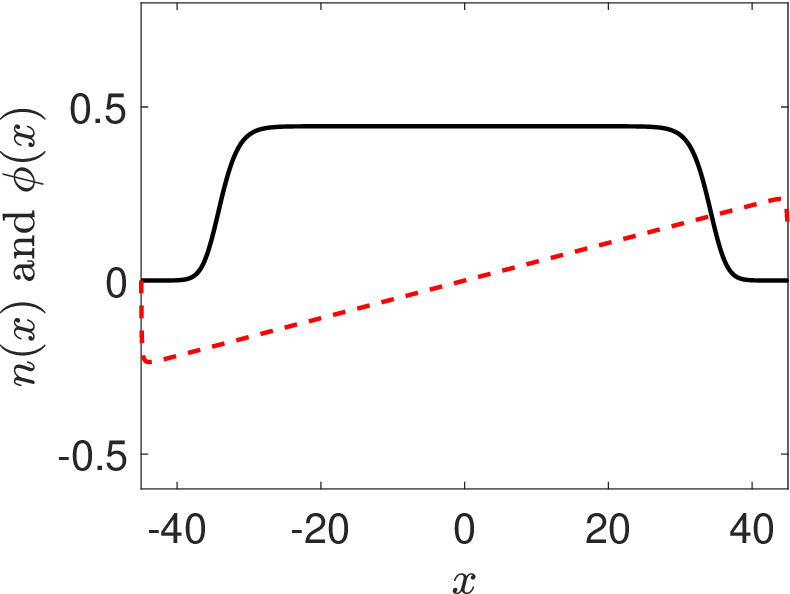}
\includegraphics[width=5.cm,height=3.cm,angle=-0]{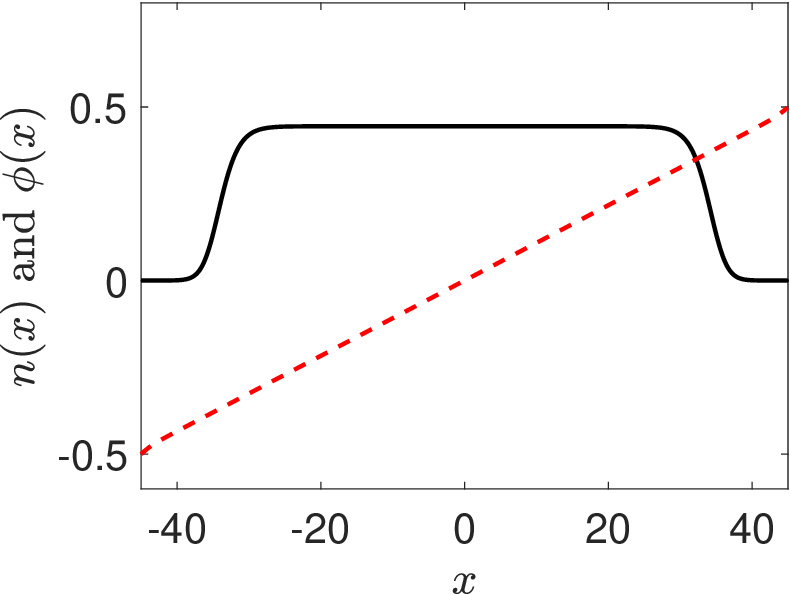}
\includegraphics[width=5.5cm,height=3.7cm,angle=-0]{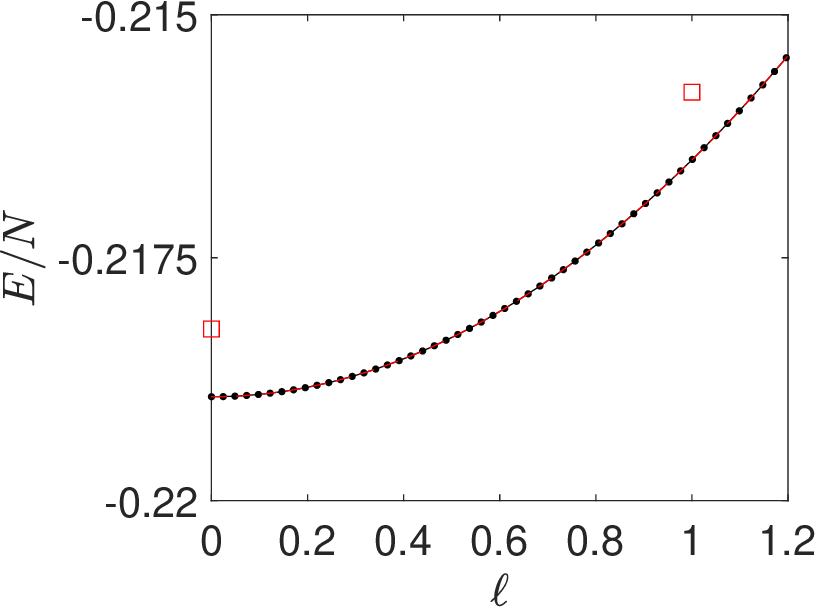}
\caption{Data for ``medium" droplet density, with $N = 30$ and $R = 45/\pi$,
i.e., $n_0 = 1/3$; middle, right dot in the phase diagram of Fig.\,1. The 
density $n(x)$ (solid curve) and the phase $\phi(x)$ divided by $2 \pi$ 
(dashed curve) of the order parameter $\Psi(x) = \sqrt{n(x)} e^{i \phi(x)}$, 
for (from top to bottom) $\ell = 0, 0.4883$ and 0.9766. The bottom plot 
shows the dispersion relation. The two square points at $\ell = 0$ and 
$\ell = 1$ show the energy of the homogeneous solution. The density is 
measured in units of $|\Phi_0|^2$, $N$ in units of $N_0$, the length in 
units of $x_0$, the energy is measured in units of $e_0$, and the angular 
momentum in units of $\hbar$.}
\end{figure}

\section{Connection with solitary-wave excitation}

The yrast states that we have evaluated in the previous section minimize the
energy functional of Eq.\,(\ref{tdcccc}), for a fixed expectation value of the
angular momentum and a fixed atom number. The last term in the energy functional
$- \Omega \langle {\hat \ell} \rangle$ may be written trivially as $-(\Omega/R) 
\langle {\hat p} \rangle$, where ${\hat p}$ is the operator of the linear momentum. 

If one looks for travelling-wave solutions, i.e., solitary-wave solutions of the
form $\Phi(x-ut)$, where $u$ is the propagation velocity of the wave, the time 
derivative in Eq.\,(\ref{1steq}), $i {\partial \Phi}/{\partial t}$, becomes 
$- i u {\partial \Phi}/{\partial x}$, i.e., $u {\hat p} \Phi$. Therefore, there 
will be a term in the energy functional which is $- u \langle {\hat p} \rangle$, 
i.e., one ends up with the same energy functional as that of Eq.\,(\ref{tdcccc}), 
with $u = \Omega R$ \cite{SMKJ}. We thus conclude that the yrast states are 
solitary-wave solutions. In addition, since we minimize the energy functional of
Eq.\,(\ref{tdcccc}) with respect to $\ell$, the velocity of propagation $u =
\Omega R$ is simply given by the slope of the dispersion relation, i.e., $u = 
R \partial [E(\ell)/N]/\partial \ell$. 

Given the above, we may derive various results regarding the properties of our 
(solitary-wave) solutions \cite{Carr1, Carr2, SMKJ}, as we describe below.

\subsection{Case of ``low" density}

Starting with the case of low densities, Fig.\,3, the attractive nonlinear 
term is the dominant one and the problem resembles that of a scalar condensate 
with an attractive effective interaction \cite{Ueda, GMK}. Here, we have a localized 
droplet, which rotates around the ring, essentially executing classical motion. 
The velocity of propagation is $u = \ell/R$, which is the classical expression.  

\subsection{Case of ``high" density}

Turning to the case of ``high" densities, Fig.\,4, the dominant nonlinear term is 
the repulsive one and the problem resembles that of a scalar condensate with a 
repulsive effective interaction. 
 
For low values of $\ell$ we have sound waves. As $\ell$ increases, both the 
propagation velocity, as well as the minimum of the waves decrease. For $\ell 
= 1/2$ we get a ``dark" wave, i.e., the minimum of the density vanishes, and 
so does the velocity of propagation, $u$. For a ring of a finite length, the 
static wave ($u=0$) occurs for a value of $\ell > 1/2$, while the ``dark" 
wave still occurs for $\ell = 1/2$. Therefore, the solitary wave in this 
case cannot be static and dark at the same time \cite{SMKJ}.    

In the thermodynamic limit of a ``large" ring, $L = 2 \pi R \to \infty$, and 
a large number of atoms, $N \to \infty$, with a finite ratio $N/L$, one gets 
the ``traditional" solitary-wave solutions, at least qualitatively see, e.g., 
Refs.\,\cite{Carr1, SMKJ}.

\subsection{Case of ``medium" density}

The final case of medium densities is the most interesting. Here, the attractive 
and the repulsive nonlinear terms are of comparable size. For small values of
$R$ (Fig.\,5, with $N = 20/3$ and $L = 20$), the density of the solitary wave 
may vanish, however the propagation velocity does not, as in the case of solitary 
waves in scalar condensates, in ``small" rings \cite{SMKJ}. For large values of 
$R$ (Fig.\,8, with $N = 30$ and $L = 90$) we have the trivial case of center-of-mass 
excitation, where the droplet rotates around the ring, without any change in its
shape, for all values of $\ell$. 

For intermediate values of $R$, the solutions that we have found do not have an 
analogue with the ones on scalar condensates. Starting with the solutions shown 
in Fig.\,6 (with $N = 15$ and $L = 45$) the interesting feature here is that 
there is a ``hybrid" solution, where for low values of $\ell$ we have the usual 
sound waves and the dispersion relation has a negative curvature. For higher 
values of $\ell$ the curvature changes sign and it becomes positive, as in the 
phonon-roton spectrum \cite{PS}. As a result, the velocity of propagation $u$ 
of the solitary wave is a decreasing function of $\ell$ for low values of $\ell$, 
however as $\ell$ increases, it starts to increase, i.e., it is not a monotonic 
function of $\ell$.

Turning to Fig.\,7 (with $N = 15$ and $L = 45$), we see that the yrast state 
for low values of $\ell$ corresponds to sound waves. At some finite value of 
$\ell$, we have a level crossing and a discontinuous phase transition to the 
other family of solutions, where we have solid-body-like motion. At this point 
the velocity of propagation changes discontinuously. A somewhat similar curve 
has been derived in Ref.\,\cite{POL}, however according to our results, the 
curve that corresponds to sound waves crosses the other one and continues for 
some larger range of $\ell$. 

\section{Summary and overview}

In this study we investigated the rotational properties of a quantum droplet 
which is confined in a ring potential. We worked at fixed angular momentum 
and determined the lowest-energy state of the droplet for various values of 
the angular momentum. As we argued, the derived solutions are also 
solitary-wave solutions.

In free space quantum droplets are self-bound, due to the balance between the 
two nonlinear terms, which have opposite sign. In the case of a ring potential 
that we consider here, depending on the atom number and the radius of the ring, 
we may tune the density and through that we may also tune the relative magnitude 
of these terms. In addition, we have the periodicity of the problem. This affects 
not only the density, but also the phase of the order parameter, which is allowed 
to differ by multiples of $2 \pi$ around the circumference of the ring.

The combined effect of the nonlinearity and of the periodic boundary conditions 
gives rise to the effects that we have presented in this study. The most 
interesting regime is the one where the two nonlinear terms have a comparable
size and also the size of the droplet is comparable with the circumference of 
the ring.

We have seen a class of solutions which resemble solid-body motion, where the 
spatially localized droplet rotates around the ring. In the other class of 
solutions we have density depressions, as in the ``classical" solitary-wave 
solutions with an effective repulsive interaction. These two classes of states 
may coexist, and we have even seen level crossings between them as the angular
momentum is varied, which will give rise to discontinuous phase transitions,
if one follows the yrast state. Interestingly enough, we have also seen ``hybrid" 
solutions, where the corresponding dispersion relation resembles that of the 
phonon-roton spectrum.

From the above it is clear that a rotating quantum droplet in a (quasi) 
one-dimensional ring is an interesting superfluid system. With the remarkable 
progress that we have seen on the experiments on quantum droplets recently, 
it would be interesting to confirm these results experimentally. Furthermore, we 
have neglected in the present study the effect of the ``width" of the ring potential. 
If one works with a torus, or an annulus, the problem will have an even more rich 
structure. This is worth investigating in the future, both theoretically, as well
as experimentally. 

Last but not least, in the present study we focused on the symmetric
case. Clearly, it would be interesting to consider also the asymmetric problem. 
In this case we no longer have one order parameter and a single equation, but rather 
two order parameters and two coupled nonlinear equations. Furthermore, at least when 
the population imbalance is not very small, we still have the formation of a droplet,
while the extra atoms of the majority component surround the droplet. As a result,
the two problems are rather different. Various questions arise immediately, including 
how the angular momentum is carried, what is the form of the dispersion relation in 
this case, etc. Reference \cite{Steffi} has investigated this problem and has derived 
various interesting results, which, however, differ from the ones presented in this 
study.

\end{document}